 \documentclass[final,5p,times,twocolumn]{elsarticle}

\usepackage{amssymb}
\usepackage{amsmath}
\usepackage{float}

\usepackage{graphicx}  
\usepackage{dcolumn}   
\usepackage{bm}        
\usepackage[T1]{fontenc}
\usepackage[utf8]{inputenc}
\usepackage{CJKutf8}

\usepackage{mathtools}

\DeclarePairedDelimiter\floor{\lfloor}{\rfloor}

\biboptions{square, sort&compress}

\journal{International Journal of Mass Spectrometry}

\begin{document}

\begin{frontmatter}



\title{Improving wide-band mass measurements in a multi-reflection time-of-flight mass spectrograph by usage of a concomitant measurement scheme}


\author[WNSC]{P.~Schury}
\author[RIKEN]{Y.~Ito}
\author[RIKEN]{M.~Rosenbusch}
\author[WNSC]{H.~Miyatake}
\author[RIKEN,WNSC]{M.~Wada} 
\author[NMSU]{H.~Wollnik}

\address[WNSC]{Wako Nuclear Science Center (WNSC), Institute of Particle and Nuclear Studies (IPNS), High Energy Accelerator Research Organization (KEK), Wako, Saitama 351-0198, Japan}
\address[RIKEN]{RIKEN Nishina Center for Accelerator Physics, Wako, Saitama 351-0198 Japan}
\address[NMSU]{New Mexico State University, Department of Chemistry and BioChemistry, Las Cruces, NM, USA}

\begin{abstract}
\newline

We introduce a new concomitant referencing mode for operating a multi-reflection time-of-flight mass spectrograph (MRTOF-MS), wherein the reference and analyte ions are interleaved on a cycle by cycle bases.  Using this mode, we demonstrate an improved technique for performing wide bandwidth mass measurements via MRTOF-MS.  This new technique offers a simplified analysis and high-accuracy.  

\end{abstract}


\end{frontmatter}


\section{Introduction}
\label{secIntro}

\par The multi-relfection time-of-flight mass spectrograph (MRTOF-MS), first proposed more than 20 years ago \cite{WollnikMRTOF}, is rapidly gaining favor at online radioactive ion (RI) beam facilities, both as an isobar separator and for performing mass measurements \cite{CaribuMRTOF, TITAN-MRTOF, PILGRIM, SlowSHE, KreimCaNature}.  These devices are capable of mass resolving powers exceeding~10$^5$ and measurement times on the order of milliseconds \cite{SlowSHE}, while having an improved immunity, as compared to Penning traps, to systematic errors induced by contaminants~\cite{Schury2014}.

\par While isotope separation on-line (ISOL) facilities can typically deliver low-energy beams of ions within a single isobar chain to an MRTOF-MS, when such devices are connected to a gas cell \cite{SlowSHE-GC}, for use with fusion-evaporation products or in-flight fragmentation beams, a natural consequence is the delivery of numerous isobar chains.  Such cocktail beams provide an opportunity for the MRTOF-MS to efficiently utilize online resources by analyzing a great many RI simultaneously.  However, as we have previously described~\cite{Previous, Schury2014} and will reiterate below, the multi-reflection nature of the MRTOF-MS results in a non-intuitive $m/q$ reordering beyond a certain mass bandwidth.  

\par The ability to accurately mass analyze the contents of such cocktail beams as produced by in-flight fragmentation, especially in a simple manner that allows ardent reviewers to recalculate masses from reported data, is highly desirable.  As one such example, isotopes believed to participate in the astrophysical r-process are produced at fairly low rates at even the most powerful facilities.  However, when produced by in-flight fission and fragmentation, numerous such isotopes can be delivered simultaneously.  A technique that could simultaneously analyze the entirety (or even majority) of such a cocktail could reduce the time required to study a broad mass region by an order of magnitude or more.

\par We have previously reported \cite{Previous} a method to analyze such large mass bandwidth cocktails.  Since then, we have made substantive modifications to the apparatus which greatly improve the reliability of mass measurements wherein the reference and analyte make different numbers of reflections in the MRTOF-MS.  These include (see Fig.~\ref{figApparatus}) an improved ion trap configuration, the addition of a focusing element and pair of dual ion steerers to improve alignment of the ion beam with the optical axis of the MRTOF-MS, and the implementation of a modified timing system to allow rapid interleaving of reference and analyte ions.

\par While our earlier technique was useful at identification of ions in a large mass bandwidth, it was also complicated and required at least two measurements be performed in series under substantially similar conditions, making it prone to errors from {\it e.g.}~slight voltage drifts.  Utilizing our presently unique ion preparation trap geometry, we have been able to modify the operation of the MRTOF-MS to allow two measurements to be made nearly in parallel, a scheme we call ``concomitant referencing".  This in turn has allowed for development of an improved method to determine the masses of ions within a large mass bandwidth. 

\par This new method bootstraps on the previously developed method.  The original method is utilized to determine the $m/q$ and the number of laps analyte ions make in the MRTOF-MS, with an accuracy of a few parts per million in $m/q$.  Once the $m/q$ and number of laps are known, our new method can determine the $m/q$ more precisely -- with a relative mass accuracy of $\sim$10$^{-7}$.  This new operational mode and improved analytical framework makes MRTOF-MS mass spectrometry highly-competitive with storage rings \cite{ESR, CSR, RRR} in terms of mass accuracy and bandwidth.

\section{Apparatus}
\label{secApparatus}

\par
\begin{figure}[H]
 \includegraphics[width=0.48\textwidth]{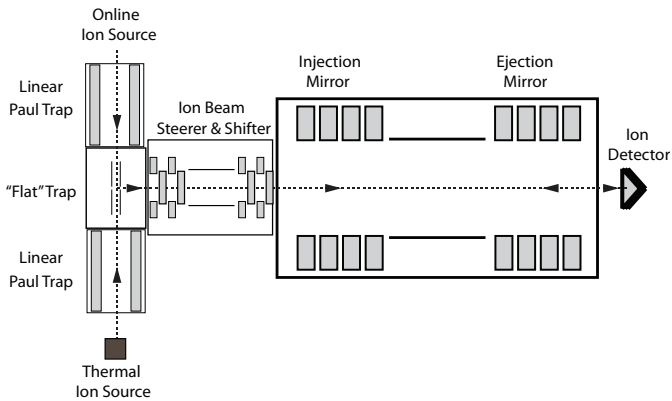}
\caption{Sketch of the MRTOF-MS system.  Reference and analyte ions accumulate in the linear Paul traps, and are transferred to the flat trap in an alternating fashion.  After cooling in the flat trap, ions are orthogonally ejected toward the MRTOF-MS.  A pair of dual ion beam steerers with a focusing element in-between allows the beam to be well-aligned with the optical axis of the MRTOF-MS.}
\label{figApparatus}
\end{figure}

 Our implementation of the MRTOF-MS (see Fig.~\ref{figApparatus}) uses a pair of electrostatic ion mirrors, with a single refocusing lens and a long field-free drift region between the mirrors.  Fast high-voltage switches are used to lower the potentials applied to the outermost electrodes of each electrostatic mirror in order to allow ions to enter and exit the MRTOF-MS.  The device is described in great detail in Ref.~\cite{Schury2014}.  

\par To achieve optimal performance from the MRTOF-MS, ions must be injected as a brilliant ion pulse, with low energy spread, and well-aligned to the MRTOF-MS optical axis.  To achieve these requirements, a suite of radio-frequency (RF) ion traps has been implemented, as shown in Figs.~\ref{figApparatus}~and~\ref{figTrapPotentials}a.  These RF ion traps are pressurized with $\sim$10$^{-2}$~mbar helium in order to cool ions.  At the heart of this ion trap suite is the ``flat" trap \cite{FlatTrap} -- a linear Paul trap constructed in a flat geometry.  Linear Paul traps are installed on both sides of the flat trap, to accumulate and store ions prior to their transfer to the flat trap.  A thermal ion source from HeatWave Labs, installed behind one of the linear Paul traps, provides offline reference ions, primarily Na$^+$, K$^+$, Rb$^+$, and Cs$^+$.

\par The linear Paul traps and the flat trap are each constructed using printed circuit boards (PCBs) as shown in Fig.~\ref{figTrapPotentials}a.  The linear Paul traps were each built from four PCBs arranged in a box configuration. The circuit boards are 135~mm by 8~mm with 15 segments connected by thin film SMD resistors and capacitors on the back side.  By capacitively coupling RF signals, with a 180$^\circ$ phase shift between adjacent PCBs, an RF psuedopotential is produced for radial ion confinement.   To allow axial manipulation of ions, DC voltages can be applied at the two outermost electrodes as well as the fourth electrode from the flat trap side.  These three DC voltages can be rapidly and simultaneously switched to produce either a DC potential well to axially confine ions or a monotonous axial gradient to transfer ions to the flat trap (see Fig.~\ref{figTrapPotentials}b-e).  Reference ions accumulate continuously in one linear Paul trap, while analyte ions continuously accumulate in the other linear Paul trap.

\par  The flat trap is composed of a pair of PCBs mounted on an aluminum support frame.  Each of the PCBs have three electrode strips, with the center strip being divided into 7 segments (see Fig.~\ref{figTrapPotentials}(a)).  The flat trap operates in so-called unbalanced mode, where a single-phase RF signal is applied to the outer strips of each PCB to produce a confining radial pseudopotential, while only DC voltages are applied to the segments of the central strip in order to produce an axial potential well.  There is a 0.8~mm diameter hole in the middle of the centermost electrode of each PCB; using a pair of switches to rapidly change the voltage applied to the centermost electrode of each PCB, a dipole electric field orthogonal to the plane of the flat trap can be produced to eject ions orthogonally.  This capability provides a simple means to accept two separate sources of ions, as well as presenting the opportunity to eject ions in either of two directions.

\par Between the flat trap and the MRTOF-MS is a pair of dual ion steerers (``Ion Beam Steerer and Shifter" in Fig.~\ref{figApparatus}) with an ersatz lens in-between them.  The recently added second dual steerer and focusing element have improved the alignment of the ion pulse ejected from the flat trap with the optical axis of the MRTOF-MS.  

\par The proper operation of the MRTOF-MS relies on an in-house designed, FPGA-based timing system.  This timing system provides timing signals for changing the voltage configurations of the traps (accumulation, cooling, ejection), the configuration of the MRTOF-MS mirrors (injection, storage, ejection), and ejection from the flat trap, as well as setting the RF phase of the flat trap at the moment of ejection.

\par The times-of-flight of ions passing through the MRTOF-MS are measured with the use of an MCS6A multi-stop time-to-digital converter (TDC) from FAST ComTec.  The flat trap ejection triggering signal also serves as the TDC start signal.  A MagneToF ion detector from ETP, installed downstream from the MRTOF-MS, provides TDC stop signals when ions strike it after leaving the MRTOF-MS.

\par The typical measurement cycle for the MRTOF-MS proceeds as follows.  First, ions accumulate and cool in the flat trap.  Before ions are ejected from the flat trap, the voltage applied to the outermost electrode of the injection-side mirror is reduced to allow ions to pass.  In order to give the high-voltage switch sufficient time to settle, this action is perfomed several microseconds before the ejection of ions from the flat trap.  Based on predetermined time-of-flight parameters, the trapping voltage is restored to the injection-side mirror when the ions under analysis are near the turning point of the ejection-side mirror.  The ions then reflect between the mirrors for a duration chosen such that ions of a specific $A/q$ undergo a predetermined number of reflections.  After the ions have undergone the desired number of reflections, and while they are near the turning point of the injection-side mirror, the voltage applied to the outermost electrode of the ejection-side mirror is reduced so as to allow the ions to pass.  The ions will then travel to the ion detector and produce the TDC stop signals.

\section{Concomitant referencing}
\label{SecConcomitant}
\par In order to determine analyte ions' masses from their times-of-flight requires time-of-flight data from reference ions of well-known mass.  Ideally, isobaric references would be available alongside the analyte ions.  This cannot be guaranteed, however.  A more general methodology, when simultaneous isobaric reference species are not available, would be to alternate between reference and analyte measurements.  Such reference measurements additionally allow correction of drifts in the ToF spectral peaks, {\emph e.g.\ }from thermal expansion/contraction and power supply voltage drifts.  However, it is only possible to correct drifts which are slow compared to the time between reference measurements.  To maximize this correction capacity it is desirable to alternate between analyte and reference as quickly as possible.  We have developed such a referencing technique, made possible by our flat trap geometry and referred to as ``concomitant referencing" \cite{CR_Patent}, wherein the reference and analyte measurements alternate on a time-scale of tens of milliseconds.

\par
\begin{figure}[H]
 \includegraphics[width=0.48\textwidth]{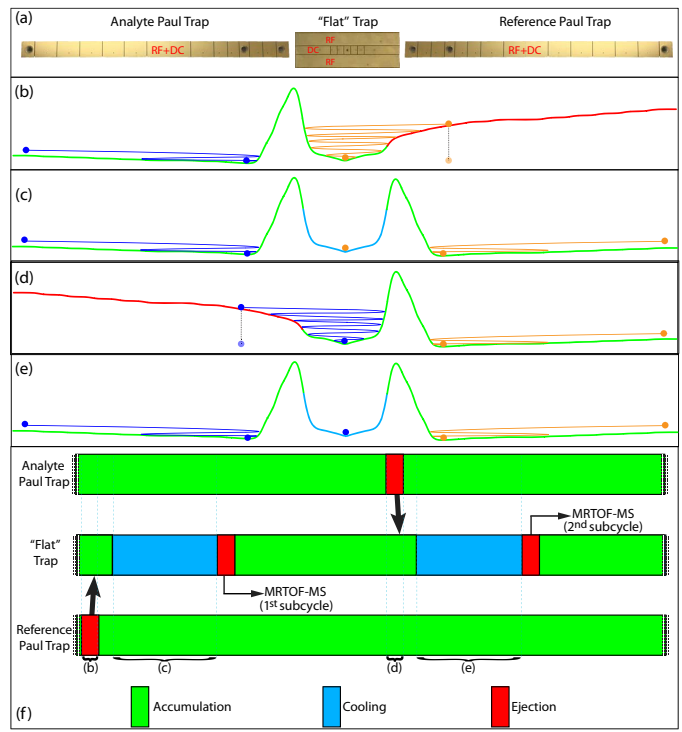}
\caption{The concomitant referencing scheme allows nearly simultaneous measurement of analyte and reference.  (a) Photographs of the PCBs used to build up the linear Paul traps and the flat trap.  Relative scale is approximately accurate.  (b)--(e) Axial potential distribution in each phase of the concomitant referencing scheme, calculated with SIMION.  (b) Reference ions are transferred from one linear Paul trap to the flat trap while analyte ions continue to accumulate in the other linear Paul trap.  (c) Reference ions cool in the flat trap while reference and analyte ions simultaneously accumulate in the linear Paul traps.  At the end of this phase, the reference ions will be ejected from the flat trap and sent to the MRTOF-MS for the first subcycle's ToF measurement. (d) Analyte ions are transferred from one linear Paul to the flat trap while reference ions accumulate in the other linear Paul trap.  (e) Analyte ions cool in the flat trap while reference and analyte ions accumulate in linear Paul traps.  At the end of this phase, the analyte ions will be ejected from the flat trap and sent to the MRTOF-MS for the second subcycle's TOF measurement.  (f) Timing arrangement in the concomitant reference scheme.  The linear Paul traps are accumulating ions for more than 90\% of the cycle duration.  The color-coding of the axial potential curves in (b)$\sim$(e) uses the legend in (f).}  
\label{figTrapPotentials}
\end{figure}

 The concomitant referencing scheme, detailed in Fig.~\ref{figTrapPotentials}, splits the measurement cycle into at least two sub-cycles, taking full advantage of the ability to transfer ions into the flat trap from two separate directions and extract them orthogonally.  Both analyte and reference ions continuously accumulate in their respective Paul traps.  At the start of one sub-cycle, the accumulated reference ions are quickly transferred to the flat trap, where they then cool until being orthogonally ejected toward the MRTOF-MS.  At the start of the next sub-cycle, the analyte ions undergo the same procedure.  By nearly constantly accumulating both analyte and reference ions in separate Paul traps, the concomitant referencing method has the benefit of an effective duty cycle exceeding 90\%.

\par  Within each subcycle the MRTOF-MS injection and ejection timings are adjusted such that analyte and reference ions experience essentially the same conditions.  The injection mirror closing time is adjusted to ensure that the ions are located in the vicinity of the ejection mirror when the injection mirror closes (thereby ensuring neither experience an energy boost from the changing electric field of the injection mirror).  The time at which the ejection mirror is opened is adjusted to ensure reference and analyte undergo the same number of laps in the MRTOF-MS before ejection.  Additionally, the MRTOF-MS ejection timing in each subcycle is adjusted such that the analyte and reference ions arrive at the detector at the same time relative to the switching of the ejection mirror voltage.  Although tests of the ejection switch (see Refs~\cite{Previous} and \cite{Schury2014}) indicate that the voltage stabilizes after $\sim$5~$\mu$s, this practice ensures that the reference and analyte ions experience maximally similar electric fields while leaving the MRTOF-MS.  Furthermore, because the reference and analyte measurement's spectra can be separately ``tagged" in the TDC data acquisition system, their analysis can be decoupled (reference ions will never eclipse analyte ions and vice versa).

\par More importantly, however, this method greatly improves the precision and accuracy of the mass analysis.  With a typical cycle time of 30~ms and a nominal mass resolving power of $R_\textrm{m}$=150\,000, if the reference ion source supplies the MRTOF-MS with one ion per cycle then a one~part-per-million (ppm) precision reference can be produced every 1.5~s.  This allows for the correction of time-of-flight drifts, resulting from voltage fluctuations and thermal expansion of the MRTOF-MS, on the level of 1~ppm/s.  Such corrections suppress artificial peak broadening and thereby increase the precision of the mass analysis by decreasing the spectral peak width and increasing the effective mass resolving power.  As the reference and analyte ions are swapped every 15~ms, for all practical purposes they experience identical conditions with regard to sources of ToF drift.  This results in a substantial improvement in the accuracy of our measurements by removing any uncertainty in the behavior of the electric fields during the analyte measurement such as discussed in Ref~\cite{ItoLi8}.

\section{Wide bandwidth mass analysis}
\label{secWideband}
\par As we have described previously, ions of significantly different $A/q$ reflecting inside the MRTOF-MS for the same duration will undergo different numbers of laps, resulting in the observed spectral peaks not being ordered by $A/q$.  This can complicate the analysis.  We previously reported on a method \cite{Previous} to identify ions that have made a different number of laps in the MRTOF-MS than the reference ion has made.  At that time, we could confidently determine the ion's mass to the level of a few ppm.  By implementating the concomitant referencing method, however, we have been able to greatly improve upon our previous results.

\par If in addition to the reference measurement, we also know the circulation time ($i.e.$ the time required for making one lap in the MRTOF-MS) of the reference ion, we can easily calculate the mass of the analyte ion.  We do this by first determining the difference in the number of laps made by the reference and analyte:
\begin{equation}
\Delta n = n_\textrm{ref} - n_\textrm{analyte} = \floor*{\bigg{(}{t_\textrm{ref} - t_\textrm{analyte}\sqrt{\frac{m_\textrm{ref}}{m_\textrm{analyte}}}}\bigg{)}/T_\textrm{ref}},
\end{equation}
where $n$, $t$, and $m$ are the number of laps, times-of-flight, and mass-to-charge ratio, respectively, of the reference and analyte ions, and $T_\textrm{ref}$ is the circulation time of the reference ion.  Even an integer (or integer ratio, for multiply-charged ions) approximation of $m_\textrm{analyte}$ is sufficient to determine $\Delta n$.  Next we calculate the time-of-flight the reference would have had if it made the same number of laps as the analyte:
\begin{equation}
t_\textrm{ref}' = t_\textrm{ref} - \Delta n\cdot T_\textrm{ref}.
\end{equation}
Then, $t_\textrm{ref}'$ can be used in the single-reference analysis methodology \cite{ItoLi8} to determine the analyte ion's mass-to-charge ratio:
\begin{equation}
m_\textrm{analyte} = m_\textrm{ref}\bigg(\frac{t_\textrm{analyte} - t_0}{t_\textrm{ref}'-t_0}\bigg)^2,
\end{equation}
where $t_0$ is an inherent delay between the ions leaving the flat trap and the start of the TDC, previously \cite{SlowSHE} determined to be $t_0$=45(5)~ns.

\par In principle, identification of unknown analyte ions would still require the complicated method we introduced in our previous manuscript on wide bandwidth mass analysis.  However, the authors' having a primary interest in precisely determining the masses of as-yet poorly studied radioactive atomic ions, in the course of this manuscript we will focus on the ability of the simple method above (Eq.~1--3) to be used in the context of concomitant reference measurements to precisely and accurately determine the mass of analyte ions -- of known identity -- making a different number of laps in the MRTOF-MS than do the reference ions.

\section{Evaluation}

\par To test how much the concomitant referencing mode, in combination with the improved beam steering and focusing section, has improved the reliable determination of the circulation time, offline tests were performed.  A rubidium thermal ion source providing $^{85, 87}$Rb$^+$, along with a relatively small yield of $^{133}$Cs$^+$, was employed to investigate the advantages of the concomitant measurement technique for wide bandwidth mass measurements.  

\par Of initial interest was whether the previously observed recurring fluctuation pattern in the circulation time as a function of lap number persisted.  Using the same data set, we determined the circulation time using both concomitantly and consecutively (see Fig.~\ref{figSpectrum}a) measured data to quantify the performance improvement inherent in the concomitant measurement technique.  We then made use of the concomitantly determined circulation times to make mass determinations of analyte ions which performed different numbers of laps than the reference ions. 

\subsection{Procedure}

\par To perform this benchmark study, the MRTOF-MS was tuned to provide a time-focus near $n$=240~laps.  The ejection voltage gradient of the Paul trap on the thermal ion source side was then chosen so as to have $\approx$25\% of the stored ions pass through the flat trap and be recaptured in the Paul trap on the other side.

\par
\begin{figure}[ht]
\includegraphics[width=0.48\textwidth]{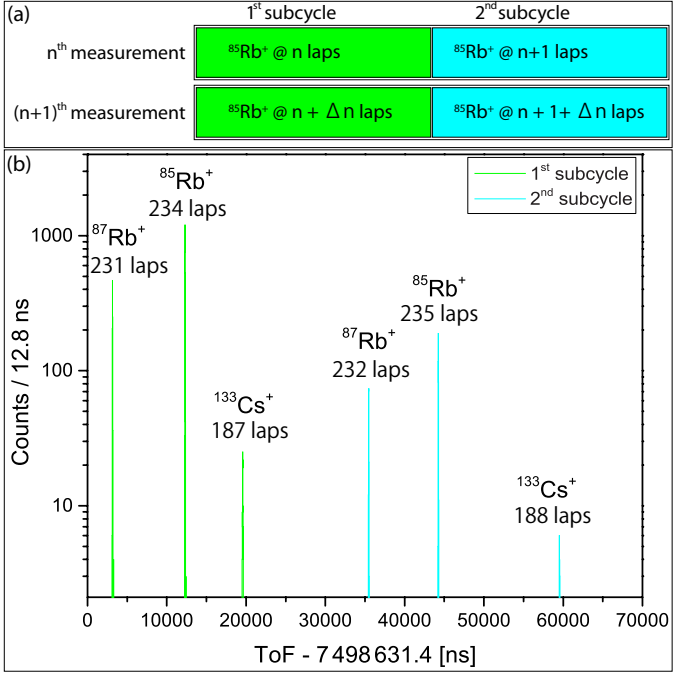}
\caption{(a) Schematic overview of the measurement method employed in this study.  Each subcycle of the $n^\textrm{th}$ and $(n+1)^\textrm{th}$ measurements together constitute a consecutive measurement, while the two subcycles of each measurement constitute a concomitant measurement.  See text for details.  (b) Spectrum observed within the measurement for $n$=234.  Note the fact that $^{85}$Rb$^+$ and $^{87}$Rb$^+$ are out of order.}

\label{figSpectrum}
\end{figure}

 At present, our concomitant measurement scheme divides a measurement cycle into two subcycles as shown in Fig.~\ref{figTrapPotentials}.  Thus, in the determination of the mass of $^{87}$Rb using $^{85}$Rb as a reference, $^{85}$Rb$^+$ would undergo $n$~laps in the first subcycle while $^{87}$Rb$^+$ would undergo $n$~laps in the second subcycle.  For this study we modified the behavior such that in the first subcycle $^{85}$Rb$^+$ would undergo $n$~laps and in the second subcycle $^{85}$Rb$^+$ would undergo $n+1$~laps, as demonstrated in Fig.~\ref{figSpectrum}a.  An example of the resulting spectrum, demonstrating the peak reordering typical of wide bandwidth mass measurements, is shown in Fig.~\ref{figSpectrum}b. 


\par Such a measurement, when performed across a wide range of $n$, should offer several advantages.  It should allow reliable determination of the reference ion circulation period $T_\mathrm{ref}$ needed for Eq.~2, and thereby will allow determination of whether there exists any inherent fluctuation in the circulation period, such as was observed in our previous study.   It will also allow a direct comparison of the previous consecutive measurement scheme with our new concomitant measurement scheme. 

\par With the MRTOF-MS timing system so configured, two series of measurements were made.  In the first measurement series, the number of laps made by $^{85}$Rb$^+$ during the first subcycle was $n(^{85}$Rb$^+)$$\in$(1, 103)~laps, while in the second series a range of $n(^{85}$Rb$^+)$$\in$(114, 330)~laps was employed.  Between consecutive measurements, the value of $n(^{85}$Rb$^+)$ was incremented by one~lap in the first series and by three~laps in the second series.  Each measurement was 180~s duration, with a 60~s wait between measurements to provide sufficient time for the voltage stabilization system to recover from changes in the noise density {\emph e.g.\ }resulting from changing the timing of the injection and ejection switches. 

\par The timing system requires that the sum of the time spent cooling in the flat trap and the time-of-flight in the MRTOF-MS must not exceed the subcycle length of 15~ms. This imposed the upper limit of $n_\textrm{max}(^{85}$Rb$^+)$$=$330~laps; longer times-of-flight would have required either reduced cooling time or longer cycle duration.  


\par For each measurement, the $^{85}$Rb$^+$ in the first subcycle was used to correct any time-of-flight drift occurring during the measurement using the procedure described in Refs~\cite{SlowSHE, Schury2014}.  After correcting for any drifts, the times-of-flight for each spectral peak corresponding to $^{85}$Rb$^+$ and $^{87}$Rb$^+$ were determined through the use of a least-square fitting algorithm utilizing an asymmetric combined Gaussian-Lorentzian function \cite{DEGH, ItoPRL, MarcoPRC} to model the spectral peak shape.  

\subsection{Analysis of circulation period}

\par The circulation period $T$ during the $n^\textrm{th}$~lap was calculated for $^{85}$Rb$^+$ using both concomitant and consecutive methodologies.  In the case of the concomitant methodology, the circulation periods were calculated as 
\begin{equation}
T_{n} = t_{n+1}^1 - t_{n}^0, 
\label{eqConcomitant}
\end{equation}
where $t$ is the time-of-flight of $^{85}$Rb$^+$ while the superscript designates the subcycle.  For mathematical simplicity, in Eqs.~\ref{eqConcomitant} and \ref{eqConsecutive} the first subcycle is assigned $i=0$ and the second subcycle is assigned $i=1$.  The flight paths always differed by one lap in the MRTOF-MS. As the spectra were accumulated in the same measurement duration, however, the ions in both spectra can be expected to experience the same prevailing conditions which would be expected to result in a minimization of fluctuations in the measured circulation period.

\par For the consecutive methodology, the circulation period was determined from each pair of consecutive measurements, wherein $^{85}$Rb$^+$ made $n$ and $n^\prime$~laps, as:
\begin{equation}
T_{n} = (t_{n+i}^i - t_{n^\prime+i}^i)/(n-n^\prime), 
\label{eqConsecutive}
\end{equation}
where the term $i$ designates the subcycle.  Each consecutive pair of measurements yielded two consecutive measurement data, one in each subcycle, differing by one lap in the MRTOF-MS.  


\begin{figure}[b!]
 \includegraphics[width=0.47\textwidth]{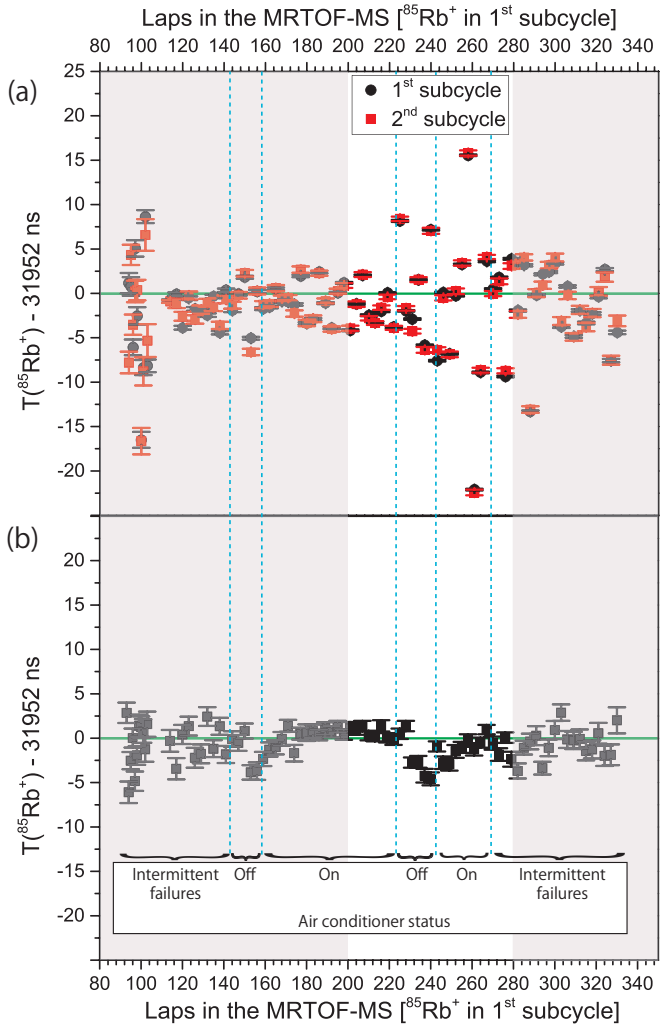}
\caption{The circulation period of $^{85}$Rb$^+$ as determined by (a) consecutive and (b) concomitant measurement analysis.  The non-highlighted range is the range of laps in which a measurement would typically be made for this tune of the MRTOF-MS.  As expected, the concomitant analysis results in greatly reduced scattering of the data.  In the concomitant analysis data, trends in the circulation time corresponding to changes in room temperature due to failures of the air conditioning unit, resulting in thermal expansion of the MRTOF-MS, can be seen.  The consecutive measurements each require two measurements.  Performing such analysis using our concomitant measurement technique yields a pair of data for each pair of measurements, one for each subcycle.  The inter-measurement scatter being much greater than the intra-measurement scatter evinces the fact that the inter-measurement scatter is a result of long-term thermal and voltage drifts of the MRTOF-MS. }  
\label{figConcomitantCirculation}
\end{figure}


\par The circulation periods, plotted as functions of the number of laps undergone by $^{85}$Rb$^+$ ions in the first subcycle, are shown in Fig.~\ref{figConcomitantCirculation}.  We limited this analysis to $n(^{85}$Rb$^+)$$\in$(90,330), as the mass resolving power fell below $R_\textrm{m}$=20\,000 for fewer than $n$=90~laps, resulting in uselessly large uncertainties which lead to relative mass uncertainties exceeding 2~ppm for the number of detected ions in these measurements.   For the data from the second measurement series, as a result of $n-n^\prime=3$, the error bars within the consecutive analysis are approximately three times smaller than those in the first measurement series.  Within the concomitant measurement analysis, the lap difference remained constant in both measurement series and the error bars reflect that.   However, it can clearly be seen that the concomitant analysis produced considerably less scattering of the data: the data from the concomitant analysis has a standard deviation of 1.7~ns and a Birge ratio \cite{Birge} of 2.2, while the data from the consecutive analysis has a standard deviation of 7.6~ns and a Birge ratio of 39.  Due to the difference in $n-n^\prime$, only a factor of 3 between consecutive and concomitant analysis Birge ratios would be expected.

\par Furthermore, the reader will notice that among the consecutively measured circulation periods in Fig.~\ref{figConcomitantCirculation}a the deviation in circulation time between the first and second subcycles within single measurements is small compared to the deviations between successive measurements.  We find that the standard deviation of the difference is  0.6~ns, a further indication that any lap-dependence of the circulation period is small, if it exists.


\par Due to the greatly reduced scattering of the data, it is possible to observe trends in the concomitantly measured circulation period which resulted as a consequence of this measurement having been performed during a particularly hot summer while the air conditioning unit was experiencing operational difficulties and failing several times per day.  Each time the air conditioning unit failed, the room quickly warmed several degrees, leading to thermal expansions. The observed excursions from the mean circulation period among measurements near $n$=150 and $n$=230~laps correlate with periods of prolonged failure of the air conditioning unit; their magnitudes are consistent with heating (and subsequent cooling) at a rate of $\approx$1$^\circ$C per minute.  The data from $n$=170 to $n$=215~laps corresponds to a prolonged period of stable operation of the air conditioning unit.     

\par The relatively larger fluctuations in the consecutive measurements can be easily understood after noting that the time-of-flight can be written
\begin{equation}
t=t_\textrm{pass}+nT 
\end{equation}
where $t_\textrm{pass}$ is the time-of-flight measured in the absence of reflections, and is typically similar to $T$ in our configuration.  If thermal expansion or voltage instabilities were to produce a change in circulation period $\Delta T$ between two consecutive measurements, the consecutive measurement analysis would yield a circulation time systematically deviated by $\frac{n}{\Delta n}\Delta T$, where $\Delta n$ is the lap difference between the two measurements.  Thus, the large excursion ($\Delta T$$\approx$20~ns) seen in the consecutive measurement data near $n$=260 can be inferred to have derived from an actual change in circulation period of $\Delta T$$\approx$0.24~ns, which is within the error bars of the concomitant measurement in this case.  



\begin{figure}[b!]
\includegraphics[width=0.48\textwidth]{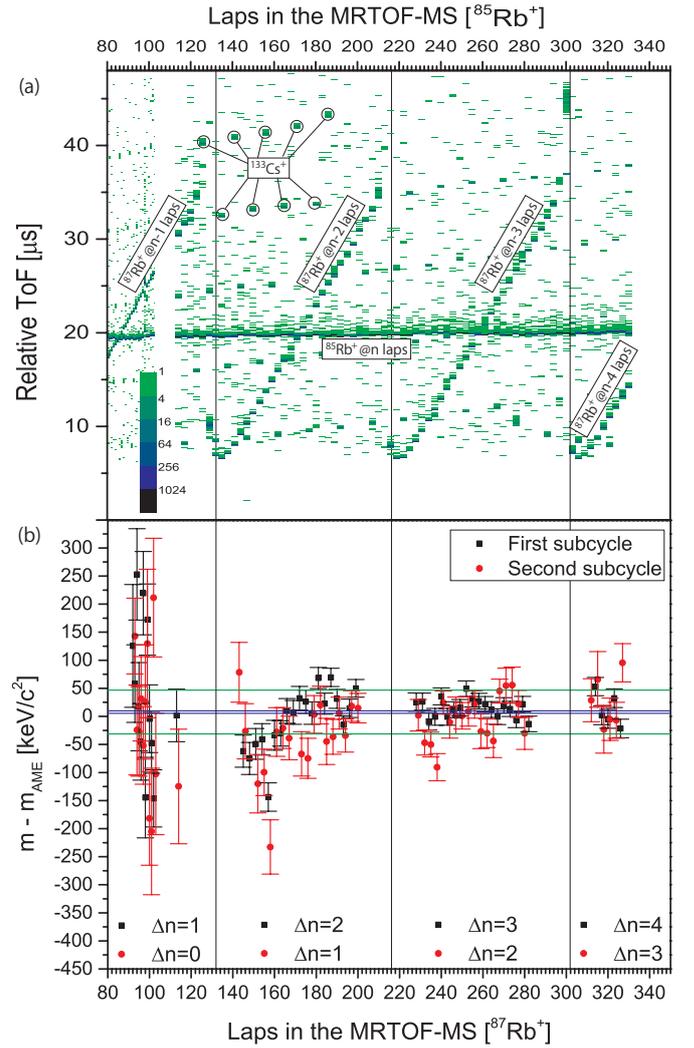}
\caption{ (a) An n-ToF color relief plot showing the spectral peak intensity as a function of number of laps and time-of-flight relative to the opening of the ejection-side mirror.  The band near 20~$\mu$s relative ToF is $^{85}$Rb$^+$, while the recurring bands titled 30$^\circ$ from the vertical are $^{87}$Rb$^+$; the recurrent points aligned at a slight angle to the horizontal are $^{133}$Cs$^+$.  As a result of $^{87}$Rb$^+$ having a longer circulation period than $^{85}$Rb$^+$, as the flight time grows longer the difference in the laps performed also increases.  When the $^{87}$Rb$^+$ is nearby or in the ejection mirror ($n\sim$130, 220, and 300) the $^{87}$Rb$^+$ ions experience a changing electric field and their times-of-flight are modified and cannot be used for mass analysis.  (b) Deviation of measured $^{87}$Rb atomic mass from AME12 values in the case of concomitant measurement analysis.  The green lines represent the weighted standard deviation bands of the data about the weighted average, while the blue lines represent the weighted average uncertainty bands.}

\label{figMassComparison}
\end{figure}

\subsection{Wide bandwidth mass analysis}

\par The two series of measurements were also used to determine the reliability of wide bandwidth mass analysis using concomitant measurements as compared to what can be achieved with consecutive measurements.  We separately analyzed spectra of each subcycle using Eqs.~1--3 with Birge ratio renormalized uncertainties for the circulation periods.  The deviations, $<$$\Delta m$$>$=$m$-$m_\textrm{AME}$, between the masses resulting from this analysis and the mass listed for $^{87}$Rb in the 2012 Atomic Mass Evaluation \cite{AME12} (AME12) can be seen in Fig.~\ref{figMassComparison}b as a function of the number of laps $^{87}$Rb$^+$ ions underwent in the MRTOF-MS.  The data are separated by the difference in numbers of laps that $^{87}$Rb$^+$ and $^{85}$Rb$^+$ travelled.  In the regions without data, $^{87}$Rb$^+$ ions were inside the ejection mirror at the moment of ejection and their flight times were significantly altered by the changing electric field inside the mirror, as shown in Fig.~\ref{figMassComparison}a.  In this analysis, full advantage was taken of the concomitant measurements by utilizing the more intense $^{85}$Rb$^+$ spectral peak in the first subcycle's spectrum as the reference for analysis of $^{87}$Rb$^+$ ions in both subcycles.  A similar analysis of the data was also made by consecutive measurement mass analysis.

\par  In the case of concomitant measurement mass analysis the weighted average deviation from the AME12 mass value of $^{87}$Rb was found to be $\Delta m$=7.7(2.2)$_\textrm{stat}$(1.4)$_\textrm{sys}$~keV/c$^2$, corresponding to a relative deviation of $\Delta m$/$m$=9.5(2.7)$_\textrm{stat}$(1.7)$_\textrm{sys}$$\times$10$^{-8}$.  For consecutive measurement analysis, the weighted average relative deviation was found to be $\Delta m$/$m$=4.3(1.2)$_\textrm{stat}$(0.2)$_\textrm{sys}$$\times$10$^{-7}$.  


\par
\begin{figure}[t]
\includegraphics[width=0.48\textwidth]{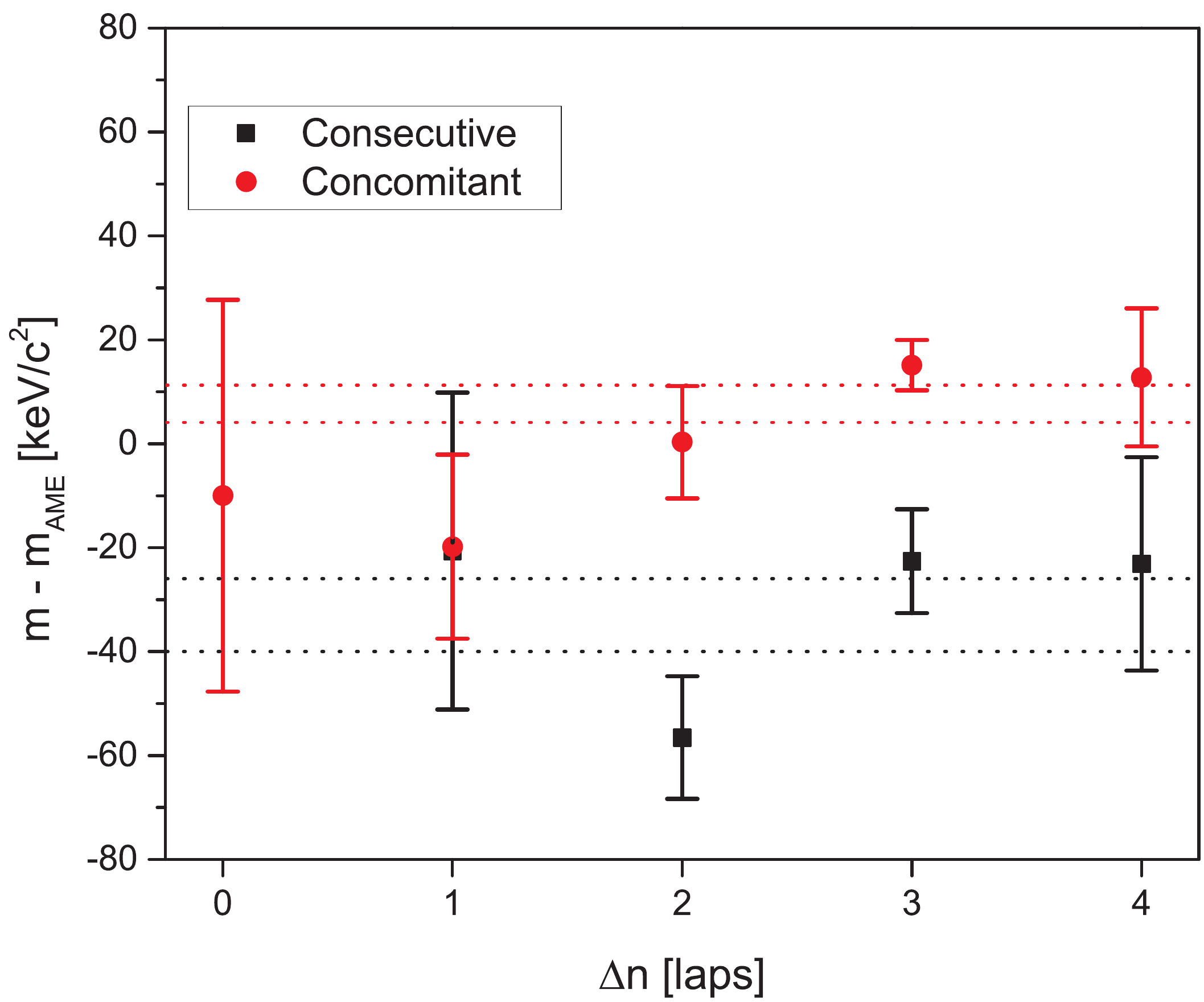}
\caption{Comparison of the accuracy and precision achieved in concomitant and consecutive measurement methodologies. The red and black dotted lines represent the 1-$\sigma$ weighted average for the concomitant and consecutive measurement methodologies, respectively.  The high precision obtained for $\Delta n$=3 is a result of that datum being derived from measurements centered near the time focus and thereby having the highest resolving power. }
\label{figMassAnalysisComparison}
\end{figure}

 The lack of any discernible lap-dependent trend in the mass deviations shown in Fig.~\ref{figMassComparison}b, even far from the time focus, suggests that lap-difference-dependent effects would dominant.  To investigate the existence of a lap-difference-dependent effect in the concomitant measurement analysis, separate $\Delta n$-based analyses of the data were made; the results are shown in Fig.~\ref{figMassAnalysisComparison}.

\par Making a best-fit line to the concomitant measurement data, a possible $\Delta n$-dependent mass deviation of $\Delta m$/$m$=5.1(1.5)$\times$10$^{-8}$/$\Delta$lap was found.  Such a dependence on the lap difference is bound to exist and indicates how well the circulation period can be determined.  If we set an accuracy goal of $\Delta m$/$m$$<$3$\times$10$^{-7}$, these results would indicate that $\Delta n$$\leq$6 would be an acceptable range for wide bandwidth measurements.  This would imply a 20\% mass bandwidth can be measured with a relative accuracy of $\Delta m$/$m$$<$3$\times$10$^{-7}$ under the presented conditions.







\section{Conclusion and Outlook}

\par We have presented a new concomitant measurement mode (as opposed to the more typical consecutive measurement mode) of operation for the MRTOF-MS, wherein reference and analyte ions are analyzed in an interlaced manner.  By alternating between measurements of the reference and analyte every 15~ms, we ensure that both species experience the same voltage and thermal fluctuations.  This operational method is only possible due to the unique flat trap geometry employed by our system.

\par By using the same ion species for reference and analyte, but making the analyte undergo one more lap in the MRTOF-MS than the reference, we could demonstrate that recent improvements in the beam alignment has removed any significant lap-dependence in the circulation period.  This removal of lap-dependence resulted in improved accuracy of mass determination, with a relative mass accuracy of $\Delta m$/$m$$\lesssim$5$\times$10$^{-7}$ being achievable in the previously reported consecutive measurements framework when analyzing $^{87}$Rb$^+$ ions making as many as four fewer laps in the MRTOF-MS than the $^{85}$Rb$^+$ reference ions.  By using the new concomitant reference framework, the achieved accuracy could be improved by a factor a five, reaching a relative accuracy of $\Delta m$/$m$=9.5(2.7)$_\textrm{stat}$(1.7)$_\textrm{sys}$$\times$10$^{-8}$.

\par The analysis was found to suffer a possible lap-dependent mass deviation of $\Delta m$/$m$=5.1(1.5)$\times$10$^{-8}$/$\Delta$lap, which would impose a 20\% mass bandwidth limit within which a relative accuracy of $\Delta m$/$m$$\leq$3$\times$10$^{-7}$ could be reliably achieved, which compares favorably to magnetic storage rings.  To improve upon this, in the near future we intend to further modify the timing system to permit an arbitrary number of measurement subcycles, allowing references measurement at several different laps to be made concomitantly.  In doing so, concomitant doublet measurements can be performed within an arbitrary mass bandwidth, while the determination of the circulation period can be made with very high precision for determination of any analyte ions outside that arbitrary mass bandwidth.


\section{Acknowledgements}
\par The authors wish to acknowledge the support of the Nishina Center for Accelerator Sciences.  This work was supported by the Japan Society for the
Promotion of Science KAKENHI (grants \#2200823, \#24224008 and \#24740142).





\bibliographystyle{elsarticle-num}



\end{document}